\documentclass[%
 reprint,
nofootinbib,
 amsmath,amssymb,
 aps,
]{revtex4-2}

\usepackage{graphicx}
\usepackage{dcolumn}
\usepackage{bm}

\newcommand{\zl}{z_{\mathrm{l}}}
\newcommand{\zs}{z_{\mathrm{s}}}
\newcommand{\rs}{r_{\mathrm{s}}}
\newcommand{\rv}{r_{\mathrm{\upsilon}}}

\newcommand{\RE}{R_{\mathrm{E}}}
\newcommand{\prs}{p_{r_{\mathrm{s}}}}

\newcommand{\wo}{W_{\mathrm{0}}}
\newcommand{\wt}{W}

\begin{document}

\preprint{APS/123-QED}

\title{Constraining compact dark matter with time-varying quasar equivalent widths}

\author{Georgios Vernardos}
    \email{Corresponding author: georgios.vernardos@lehman.cuny.edu}
\author{James Hung Hsu Chan}
\affiliation{
    Department of Physics and Astronomy, Lehman College of the City University of New York, Bronx, NY 10468, USA
}%
\affiliation{
    Department of Astrophysics, American Museum of Natural History, Central Park West and 79th Street, NY 10024, USA
}%
\affiliation{
    Institute of Physics, Laboratory of Astrophysics, \'Ecole Polytechnique F\'ed\'erale de Lausanne, Observatoire de Sauverny, 1290 Versoix, Switzerland
}%

\author{Frederic Courbin}
\affiliation{
    ICC-UB Institut de Ci\`encies del Cosmos, Universitat de Barcelona, Mart\'i Franqu\`es, 1, E-08028 Barcelona, Spain
}%
\affiliation{
    ICREA, Pg. Llu\'is Companys 23, Barcelona, E-08010, Spain
}%
\affiliation{
    Institute of Physics, Laboratory of Astrophysics, \'Ecole Polytechnique F\'ed\'erale de Lausanne, Observatoire de Sauverny, 1290 Versoix, Switzerland
}%

\date{\today}

\begin{abstract}
One of the possible explanations for dark matter is that of compact dark objects of baryonic origin, such as black holes or even planets.
Accumulating evidence, including the discovery of merging stellar mass black holes through gravitational waves, point to a population of such objects making up at least some fraction of dark matter.
We revisit a historically heavily used probe, quasar spectra, from the new perspective of time variability and gravitational lensing.
From a sample of 777 quasars selected from archival data we identify 19 that show decisive evidence of lensing by compact objects with masses measured in the range $5\times 10^{-5} < M/\mathrm{M}_{\odot} < 2\times 10^{-2}$ with 99\% confidence.
This is much lower than what is hoped to be detected by even the most futuristic gravitational wave detectors and analysis strategies, but is crucial for theories of compact dark matter, such as primordial black holes predicted from quantum phase transitions in the early Universe.
\end{abstract}

\maketitle


\section{Introduction}
The current cosmological paradigm implies that 85\% of the matter content of the Universe remains invisible to the eye~\cite{Planck2020}.
There is a plethora of possible models and interpretations for this so-called ``dark matter'' (DM), albeit still poorly constrained by observations~\cite{Bertone2018}.
If it exists at all, it is not clear if it is cold~\cite{Blumenthal1984}, or warmer, depending on the properties of a yet-to-be-discovered elementary particle.
Another possible explanation is that it consists of an undetected population of compact, too dim or dark, baryonic objects such as stars, planets, or stellar remnants like black-holes.
More exotic objects could be added to this population, such as ultra-compact DM halos~\cite{Ricotti2009,Blinov2021}, boson stars~\cite{Visinelli2021}, composite DM structures~\cite{KribsNeil2016}, and primordial black holes~\cite{Carr2024}.
By first detecting such compact objects and subsequently measuring their abundance and mass we can constrain several alternative theories for compact DM, for example, any detection below 3 M$_\odot$ would be a confirmation of the existence of non-stellar, primordial black holes~\cite{VillanuevaDomingo2021}.

\begin{figure*}
	\centering
	\includegraphics[width=0.95\textwidth]{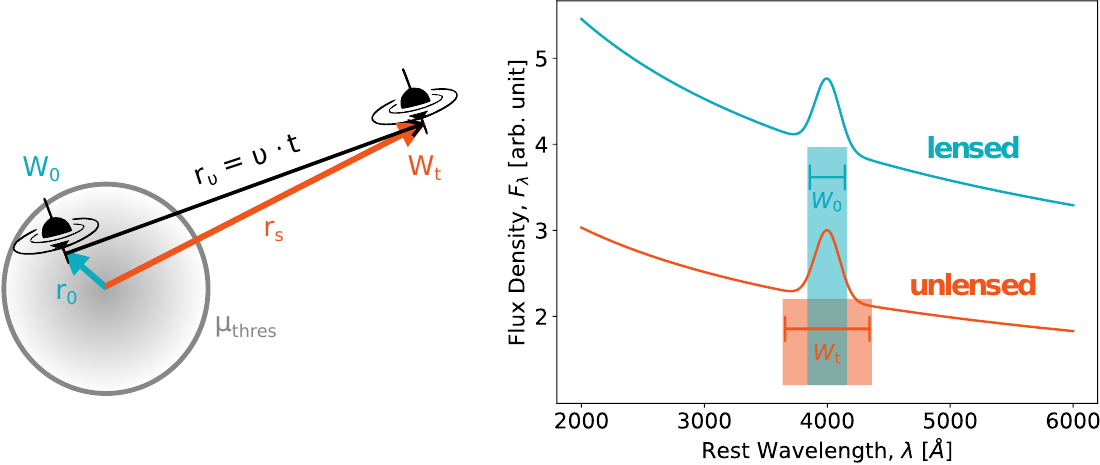}
	\caption{
		Schematic view of microlensing by an isolated compact object passing in front of a (non-strongly-lensed) quasar. {\it Left:} at some reference time, $t_0$, a quasar accretion disk (black) is located within a region of the sky (shown in grey) where the lensing magnification due to a foreground compact object is above a given threshold, $\mu_{\rm thres}$. Some time, $t$, later (many years), the quasar has moved away from this region and is not lensed anymore. {\it Right:} As the Narrow Line Region (NLR) of the quasar is much larger than the accretion disk and than the lensing cross-section, shown in black and grey on the left panel respectively, it is unaffected by microlensing. As a result, the equivalent width, $W_0$, measured at $t_0$ when the continuum was magnified with respect to the lines, becomes larger at time $t$ when there is no microlensing anymore. The present work investigates how to optimise the detection of a change in equivalent width and assesses the probability that it is due to microlensing and not intrinsic quasar variability. The time scales we consider are of the order of a decade between $t_0$ and $t$.
	}
	\label{fig:schematic}
\end{figure*}

The first place to look for compact DM has historically been the halo of our own Galaxy and one of the main probes for doing so is gravitational lensing.
Microlensing experiments in the Galactic halo (MACHO and EROS;~\cite{Alcock1993, Aubourg1993}) and bulge (OGLE;~\cite{Udalski1993, Paczynski1986}) concluded that less than 25\% of the dark matter content in the Milky Way is in the form of compact objects of $<1$ M$_{\odot}$, while more recent observations of the Large Magellanic Cloud have shown that no more than 10\% of DM can be attributed to objects with masses $< 10^3$ M$_{\odot}$~\cite{Mroz2024}.
The discovery of gravitational waves from compact binary systems~\cite{Abbott2016} with mass above 10 M$_\odot$~\cite{Hutsi2020} reinvigorated these efforts, with~\cite{Sahu2022} even reporting the discovery of an intermediate mass black hole ($\sim$ 7 M$_\odot$) from a long-duration (t$\sim$276 days) astrometric microlensing event in the Galactic bulge.
Looking for compact dark object in our Galaxy is an ongoing effort, with a new technique proposed very recently that searches for dimmed instead of amplified lensing events \cite{Bramante2025}.

Given the accepted paradigm of structure formation via hierarchical gravitational collapse, massive halos populated by galaxies are the places to look for compact DM outside our Galaxy.
Such dense places in the Universe are galaxy clusters where super-magnified stars lensed by compact objects have been detected~\cite{Oguri2018,Muller2024}.
The existence of free-floating compact objects in the inter-galactic space along random lines of sight has not been excluded either. 
Population studies of lensing effects on type I-a Supernova light curves~\cite{Zumalacarregui2018,Dhawan2023} and other transient events, such as time-delayed signals present in Gamma Ray Bursts~\cite{Paynter2021} or interference patterns in Fast Radio Bursts~\cite{Leung2022}, have also provided measurements of mass and abundance of compact objects.

But perhaps the most mainstream probe for detecting compact DM throughout the decades have been quasars.
A higher optical depth to lensing by compact objects is to be expected in strongly lensed quasars, which can display ``anomalous flux ratios'' between their lensed images~\cite{Wen2022,Dike2023}, however, similar lensing signatures can occur in field quasars if we assume the presence of a large population of almost free-floating compact objects.
The principle of the method was first proposed by~\cite{PressGunn1973} and has been continuously revisited since then~\cite{Canizares1982, Hawkins1993, Hawkins2011}.
\cite{Schneider1993} used quasar variability to place an upper limit of $\Omega \leq 0.1$ to the cosmological density of compact objects with masses $<0.1$ M$_{\odot}$, depending on the physical size of the quasar emitting region~\cite{ZackrissonBergvall2003}.
\cite{Zackrisson2003} concluded that microlensing alone was not sufficient to explain photometric variations in quasars and~\cite{Hawkins2022} postulated that a combined lensing and intrinsic variability is necessary.
\cite{Luo2020} identified 16 quasars from SDSS Stripe 82 with bell shaped light curves interpreted as candidate microlensing events.

Quasar emission regions of different physical size respond differently to lensing magnification~\cite{VernardosISSI2024}, especially the continuum and emission line regions: when a compact mass intersects the line of sight to a distant quasar but is light enough not to create resolvable multiple images ($< 10^7$ M$_{\odot}$), the small continuum region is magnified (sits entirely within the caustics), while the much larger Broad and Narrow Line Regions are barely affected.
As a result, the line equivalent width in a quasar spectrum, $W$, is smaller in presence of microlensing than in its absence - Fig.~\ref{fig:schematic} gives a schematic view of the effect. 
This method, first proposed by~\cite{Canizares1982}, was employed to study the fraction and mass of such putative compact objects~\cite{Dalcanton1994,Wiegert2003,Francis1996}.

In this work we revisit the role of spectral deformation of non-strongly-lensed quasars as a probe of compact object DM, expanding on previous works by including a time-dependence: the relative velocity between the observer, the lens, and the source can change their relative alignment and cause an observable change in the relative magnification of the continuum with respect to the emission lines.
This change is measurable through one single number: the line equivalent width (or just equivalent width, see Fig.~\ref{fig:schematic}).
Although all theoretical aspects of this analysis perspective have been well-described before, it was never considered in light of observations in the time dimension.
However, quasars are known to be intrinsically variable across wavelengths, which can lead to the same observational signature over similar (but not the same) timescales as lensing.
Therefore, we adopt a quantitative comparison of two competing effects: 1) the intrinsic continuum variability and 2) the combination of intrinsic and lensing variability.
The continuum and narrow line regions are effectively two limiting cases for lensing, i.e. a point source and an extended source large enough to be unaffected by microlensing magnification.
This allows us to use a simple analytic model for the magnification, and, to first approximation, avoid multiple lenses and extended sources.
For those quasars that favour the lensing hypothesis, we proceed by measuring the most probable lens mass.
Throughout this paper we use a standard cosmological model with $\Omega_{\rm m} = 0.3$, $\Omega_{\mathrm{\Lambda}}=0.7$, and $H_0=70$ km/s/Mpc.

\begin{figure*}
	\centering
	\includegraphics[width=0.97\textwidth]{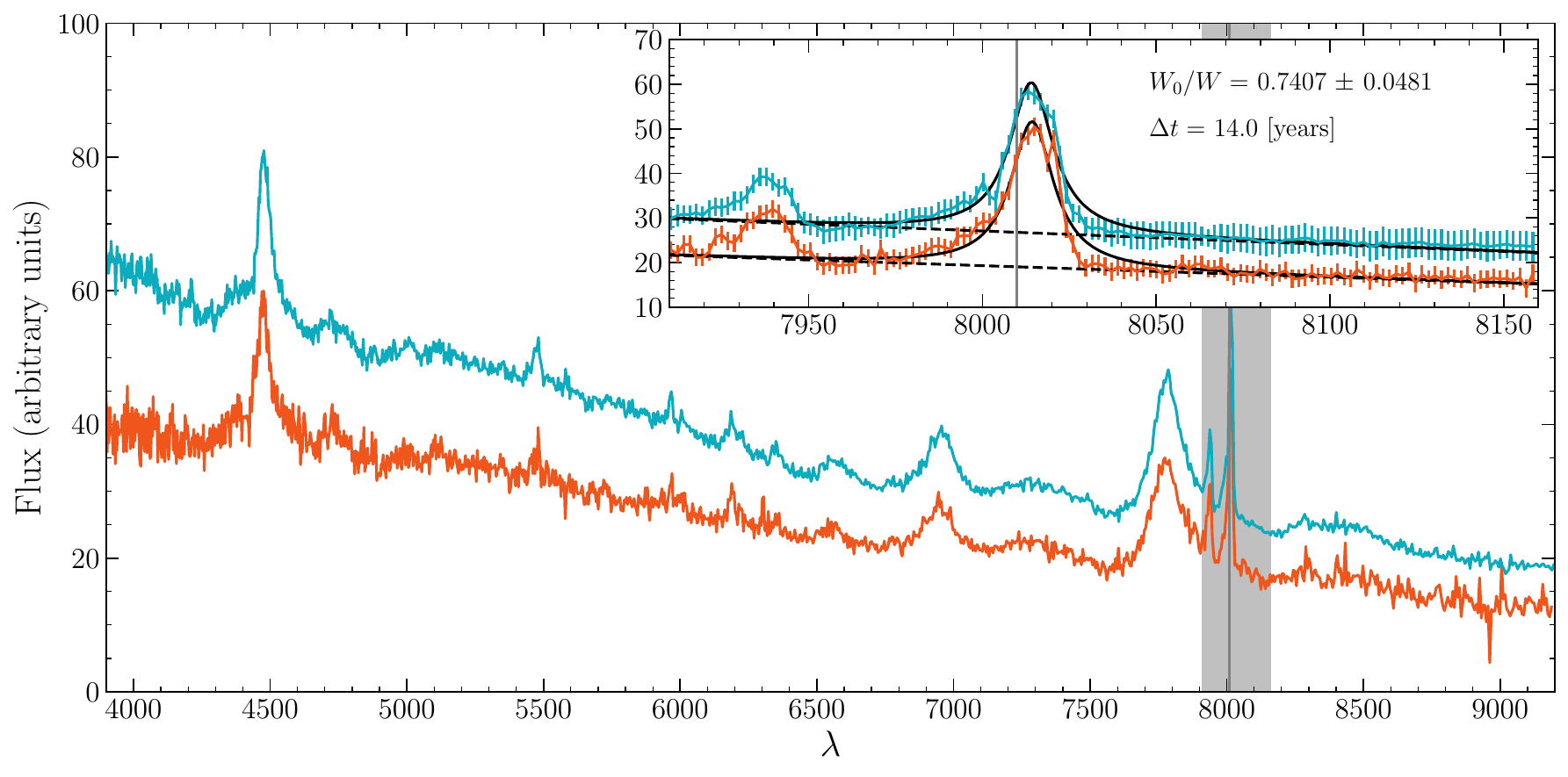}
	\caption{Example of a quasar within our sample that shows the lensing effect that is schematically described in Fig. \ref{fig:schematic}. The two spectra have been obtained by SDSS and are separated by 14 years. We fit the flux in a $\sim300$\AA~wavelength range centered on the [OIII] narrow line (vertical line), as highlighted by the grey strip and shown in more detail in the inset plot. The dashed and solid lines show the fitted continuum and total (including the emission line) flux components. Our simple model does not fit the flux everywhere, however, the inaccuracies occur consistently between the two spectra, e.g. in the wings of the line (8025-8040\AA). This leads to a quite robust measurement of the line equivalent width ratio between the spectra: $\wo/W=0.74\pm0.05$. In this quasar, the line flux varies by less than 5\% between the two spectra, meaning that the equivalent width ratio change can be attributed solely to the continuum. Errorbars are shown only in the inset plot for clarity.}
	\label{fig:spectrum}
\end{figure*}

\section{Data}
There are tens of thousands of known quasars that have been spectroscopically observed throughout the years.
Our search in archival data of the Sloan Digital Sky Survey (SDSS) resulted in 777 quasars with two spectra taken at least 7 years apart.
From these spectra, we measure the equivalent width (Eq. \ref{eq:ew_def_time}) ratios of the [OIII] (5007\AA) narrow line.
We select this line for two important reasons: 1) it always appears well within the wavelength range of the spectra for the given quasar redshifts, and 2) the corresponding emitting region of the quasar is too large to be affected by lensing from compact objects in the examined mass range ($<100$ M$_\odot$).
We emphasize that our final goal is to obtain equivalent width ratios, which are less sensitive on how we model the continuum and line flux in the spectra than individual equivalent width values.
Therefore, it is sufficient to fit each spectrum in a range of a few hundred~\AA~around the center of the emission line, using a Lorentzian profile and a straight line for the emission line and continuum flux respectively.
Our simplistic approach inevitably leads to imperfect fits and non-negligible residual flux, especially in quasars that have complex line profiles or a strong neighbouring iron line component.
Although the resulting values should be carefully used individually, their ratios are quite robust.
An example of such a fit is shown in Fig.~\ref{fig:spectrum}.

We further refine our sample of 777 quasars in order to ensure that changes in the equivalent width of the [OIII] line between two epochs originate from a change in the continuum flux based on the following criteria:
\begin{enumerate}
	\item We perform a visual inspection of the goodness-of-fit of our model to each spectrum. If the line profile is too complicated to fit or if the spectrum in general shows features beyond a simple shift in the continuum flux, then we remove the quasar from the sample.
	\item We remove quasars with measurements of equivalent width ratios with an uncertainty of more than 50\%. Such high uncertainty is a direct result of large errors in the observations.
	\item We remove quasars whose emission line flux changed by more than 10\% between the two spectra, as this would indicate some sort of intrinsic variability.
	\item We ensure that none of the other broad lines disappears between the two chosen epochs - a typical characteristic of Changing State quasars~(see fig. 1 of~\cite{Ricci2023}). We apply this criterion quite strictly and remove quasars with any significant variation of broad line flux.
\end{enumerate}
The resulting sample of 183 quasars has robust equivalent width ratio measurements, shown in the top of Fig.~\ref{fig:bayesian_factor_sample}.

\section{Time varying equivalent widths}
The equivalent width of an emission line is defined as:
\begin{equation}
	\label{eq:ew_def_time}
	W(t) = \int_{\lambda_{\rm min}}^{\lambda_{\rm max}} \frac{F_{\rm L}(\lambda)}{\mu(t)F_{\rm C}} \mathrm{d} \lambda,
\end{equation}
\noindent where $\lambda$ is the wavelength, $F_{\rm L}$ is the emission line flux profile, and $F_{\rm C}$ is the continuum flux assumed to be constant across the line wavelength range $\lambda_{\rm min} - \lambda_{\rm max}$. 
In this definition, we introduce $\mu(t)$, a generic time-dependent factor that captures changes in the ratio of continuum and emission line flux attributed to some underlying variability model.
For two equivalent width values, $\wo$ and $\wt$, separated by time $\Delta t = t - t_0$ we thus have:
\begin{equation}
	\label{eq:ew_ratio}
	\frac{\wo}{\wt} = \frac{\mu}{\mu_0},
\end{equation}
\noindent where $\mu_0$ corresponds to $t_0$ and $\mu$ resulted $\Delta t$ after $\mu_0$, i.e. $ \mu \equiv \mu(t_0+\Delta t, \mu_0)$.
The variability processes examined here are stationary, which means that they do not change by shifting them in time and thus the value of $t_0$ is irrelevant.
The value of $\mu(\Delta t,\mu_0)$ is, however, dependent on $\mu_0$ and together they form a pair of correlated random variables.
Therefore, the probability distribution of the ratio of any two equivalent width values given $\Delta t$ can be written as:
\begin{equation}
	\label{eq:ew_ratio_prob}
	p \left( \frac{\wo}{\wt} \big| \, \Delta t \right) = \int \mu_0  p_{\mathrm{\mu}} \left( \frac{\wo}{\wt} \mu_0 | \, \Delta t, \mu_0 \right) p_{\mathrm{\mu_0}}(\mu_0) \, \mathrm{d} \mu_0.
\end{equation}
\noindent This equation is the cornerstone of our work because it relates directly observable quantities with an underlying variability model.

\section{Lensing and intrinsic variability}
We employ two models in order to interpret the measured equivalent width change: 1) a purely intrinsic one (Intrinsic), based on a classic Damped Random Walk \cite{MacLeod2010}, and 2) a model that combines intrinsic variability and lensing (Combined).

For a pure intrinsic variability model, which requires the $K$-corrected absolute magnitude in the $i$-band of the quasar to be known, $M_{\mathrm{i}}$, Eq. (\ref{eq:ew_ratio_prob}) becomes (derived in detail in \cite{supplemental}):
\begin{widetext}
\begin{equation}
	\label{eq:prob_final_in}
	p^{\rm IN} \left( \frac{\wo}{\wt} \big| \, \Delta t, M_{\rm i} \right) = \int \int \frac{p_{\rm X}\left( \mathrm{log}(\frac{\wo}{\wt} \mu_0) \big| \, \Delta t,M_{\rm i},\mu_0, \mathrm{log}M_{\rm BH} \right) \, p_{\mathrm{M_{BH}}}\left( \mathrm{log}M_{\rm BH} | M_{\rm i} \right) \, p^{\rm IN}_{\mathrm{\mu_0}}(\mu_0|M_{\rm i})}{\frac{\wo}{\wt} \mathrm{ln}10} \, \mathrm{d} \mathrm{log}M_{\rm BH} \, \mathrm{d} \mu_0,
\end{equation}
\end{widetext}
where $p_{\rm X}$, $p^{\rm IN}_{\mathrm{\mu_0}}$, and $p_{\mathrm{M_{BH}}}$, the prior used to marginalize over the mass of the quasar's supermassive black hole, $M_{\rm BH}$, which is a nuisance parameter, are normal distributions with known means and standard deviations.

For a pure lensing variability model, which requires the lens mass, $M$, the quasar redshift, $\zs$, and the quasar sky coordinates, $\mathbf{\zeta}$, to be known, Eq. (\ref{eq:ew_ratio_prob}) becomes (derived in detail in \cite{supplemental}):
\begin{widetext}
\begin{equation}
	\label{eq:prob_final_le}
	p^{\rm LE} \left( \frac{\wo}{\wt} \big| \, \Delta t, \zs, \mathbf{\zeta}, M \right) = 
	\frac{2}{\pi} \int \int_{0}^{\zs} \int_{r_0-\rs}^{r_0+\rs} \frac{\mu_0  \,  \rv I_{\rm 0}\left( \frac{\rv r_{\rm c}}{\sigma^2} \right) \exp \left( - \frac{\rv^2 + r_{\rm c}^2}{2\sigma^2}\right) \, p_{\mathrm{\zl}}(\zl|\zs) \, p^{\rm LE}_{\mathrm{\mu_0}}(\mu_0)}{\left[ \left( \frac{\wo}{\wt} \mu_0 \right)^2 -1 \right]^{3/2} \sigma \, \Big\{ \big[\rv^2-(r_0 - \rs)^2\big]\big[(r_0+\rs)^2 - \rv^2\big]\Big\}^{1/2}} \, \mathrm{d} \rv \, \mathrm{d} \zl \, \mathrm{d} \mu_0,
\end{equation}
\end{widetext}
where the lens redshift, $\zl$, is a nuisance parameter that we integrate over using the prior $p_{\mathrm{\zl}}(\zl|\zs)$ described in Appendix~\ref{app:lens_redshift}, the vectors $\rv, \rs$, and $r_0$ are indicated in Fig. \ref{fig:schematic}, $I_0$ is the zero-th order modified Bessel function of the first kind, $r_{\rm c}$ and $\sigma$ have known values, and $p^{\rm LE}_{\mathrm{\mu_0}}(\mu_0)$ is the probability distribution of lensing magnification.

Our sample of quasars includes all of the parameters required by the two models: the absolute magnitude in the $i$-band, $M_{\rm i}$, the sky coordinates, $\mathbf{\zeta}$, the source redshift, $\zs$, and the date when each spectrum was obtained (required for $\Delta t$).
The only remaining parameter to set is the mass of the object acting as lens, $M$ (the density of such compact objects in the Universe plays a secondary role, as discussed below).
In the following, we explore a range of masses between $10^{-5}$ and $10^2$ M$_{\odot}$.
Obtaining Eq. (\ref{eq:ew_ratio_prob}) for the combined model is described in Appendix~\ref{app:intrinsic_and_lensing}.

\section{Model comparison}
In order to quantify which of the two models, Intrinsic or Combined, is favoured by the data, we use the Bayesian factor, $K$, described in Appendix~\ref{app:bayesian_factor}.
Figure \ref{fig:bayesian_factor_sample} shows $K$ calculated for every quasar given the measured equivalent width ratios - the higher its value the more strongly the data support the lensing hypothesis.
For small ratios the Intrinsic model is sufficient to explain the observations, but for $\wo/W < 0.65$ the Bayesian factor begins to favour the lensing hypothesis.

\begin{figure}
	\centering
	\includegraphics[width=\columnwidth]{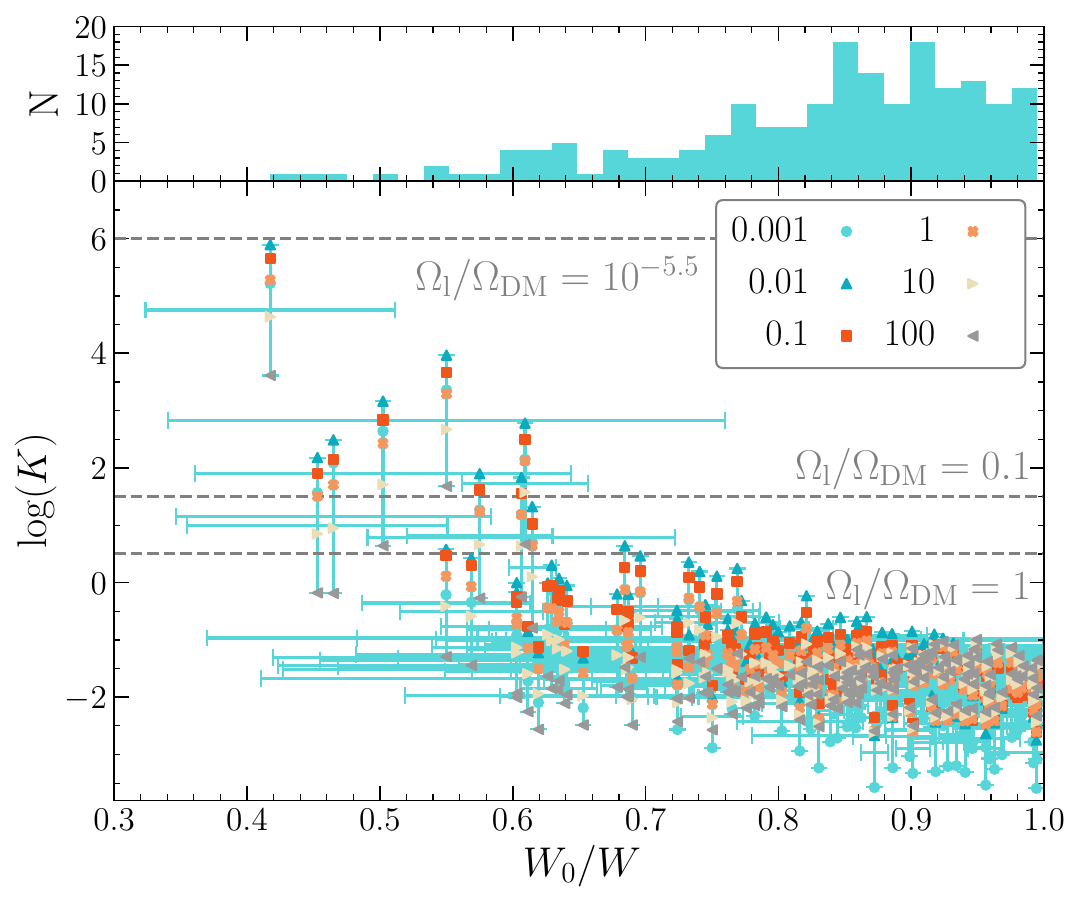}
	\caption{Comparison of the purely intrinsic and combined (intrinsic and lensing) variability hypotheses. We use the Bayesian factor, $K$, for the measured equivalent width ratio, $\wo/W$, for each of the 183 quasars in our refined sample. We plot $K$ for six different masses per quasar as indicated by the symbols, with the models for 0.01 and 100 M$_{\odot}$ (blue stars and grey triangles, respectively) having the highest and lowest $K$ per quasar. For practical purposes, we chose to place the error bar on the measurement of $\wo/W$ in the middle of the range of $K$. The solid horizontal lines denote the limit for substantial evidence ($\log K > 0.5$) in favour of the lensing hypothesis. This limit depends on the cosmological density parameter of matter in the form of compact lenses, for which we show 3 indicative values as a fraction of the dark  matter density, $\Omega_{\rm DM}$: 1, 0.1, and $10^{-5.5}$. For the latter, the density of compact objects in the Universe is too low to justify lensing in our sample of quasars under any circumstance.}
	\label{fig:bayesian_factor_sample}
\end{figure}

\section{Measuring the lens mass}
\begin{figure}
	\centering
	\includegraphics[width=\columnwidth]{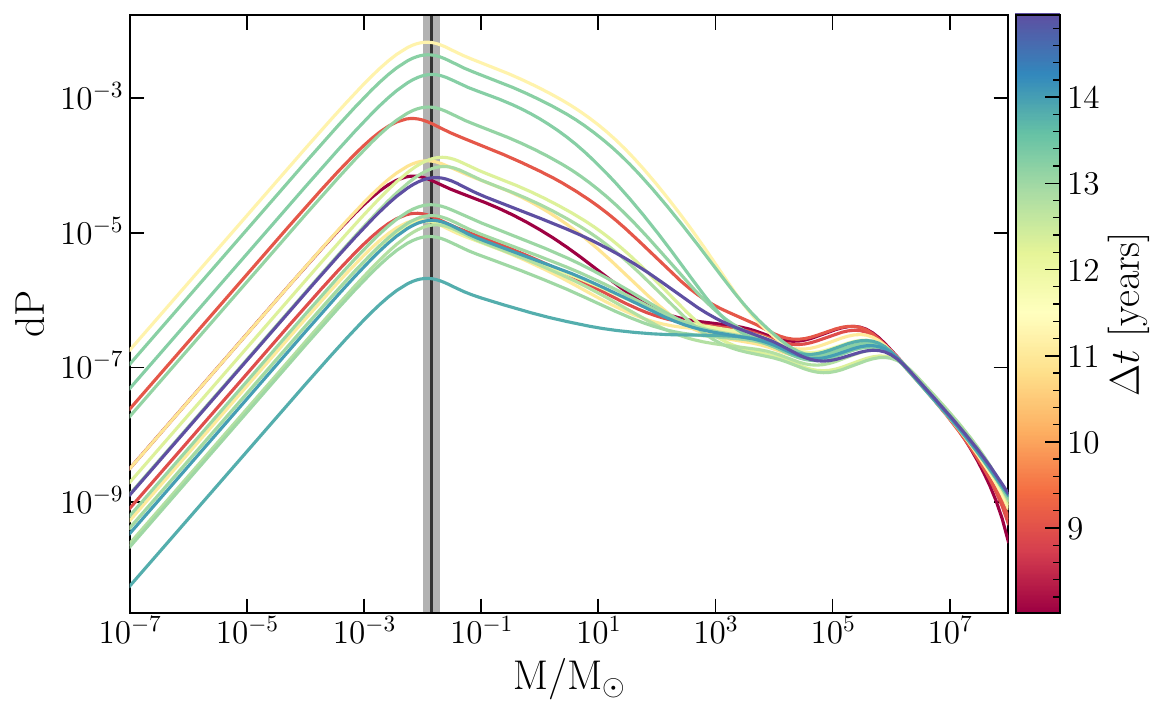}
	\caption{Posterior probability distribution of the lens mass using a uniform prior. Distributions are shown for the sub-sample of 19 quasars with the highest Bayesian factors in favour of the lensing hypothesis. The vertical line indicates the combined measurement assuming that all the lenses have the same mass and the shaded area indicates the 99\% confidence interval. The distributions are colored according to the time separation of their two spectra, $\Delta t$.}
	\label{fig:combined_fit}
\end{figure}

For a subset of our quasars, the ratio measurements are better explained by lensing variability and we can fit the Combined model to the data in order to measure the lens mass.
We select the quasars with strong evidence of lensing taking place according to the value of the Bayesian factor (e.g. $\log K > 1$ for strong evidence~\cite{Jeffreys1998}).
But this threshold depends on the density of compact objects that could act as lenses, $\Omega_{\rm l}$, and the Bayesian factor for each quasar depends on the assumed lens mass (see Fig. \ref{fig:bayesian_factor_sample}).
Hence, we select a sub-sample of quasars with $\log K > 0$ for $M=0.01$ M$_{\odot}$, the mass value with the highest $K$.
The resulting 19 quasars are then fitted using the Combined model and a choice of prior on the lens mass, as described in Appendix~\ref{app:combined_fit}, in order to obtain the posterior probability distribution of the lens mass that we show in Fig.~\ref{fig:combined_fit}.
There is a dependence on the prior, however, we can place the measured mass of a monochromatic population, i.e. all lenses having the same mass, within the range $5\times 10^{-5} < M/\mathrm{M}_{\odot} < 2\times 10^{-2}$ with 99\% confidence.

\section{Conclusions}
We revisit the idea that equivalent widths of emission lines in quasar spectra can be affected by lensing and we now examine it from the new perspective of the time dimension.
Our results show that big changes between equivalent width measurements taken sufficiently apart in time can be better explained by lensing as opposed to intrinsic variability.
We consistently find the mass of an assumed ``monochromatic'' lens population to be $\mathcal{O}(10^{-2})$ M$_{\odot}$ for quasars at various coordinates, redshifts, and with different brightness.
Our measured masses are consistent with the most probable mass for primordial black holes due to the QCD phase transition~\cite{Byrnes2018}, or possible alternatives in the 20-100 M$_{\odot}$ range~\cite{Bird2016}, and the population of $10-30$ M$_{\odot}$ black holes detected by LIGO~\cite{Abbott2016}.
Our findings are compatible with a population of free-floating planet-mass objects making up at least $10^{-5.5}$ of the total dark matter density.

While pre-selecting quasars with good SDSS spectroscopy, we enforced the following criteria: 1) the redshifts should be high enough to maximize the lensing optical depth and minimize the Einstein crossing time, 2) our chosen [OIII] narrow line would remain in the observed wavelength range out to $z_s=0.7$, and 3) available spectra in at least two epochs separated by $\sim$ 10 years, a time scale corresponding roughly to the Einstein radius crossing time of a stellar mass lens.
The sample was further refined by removing quasars displaying spectral changes that did not match a clear lensing signature, for example, Fig. \ref{fig:spectrum} and fig. 1 from~\cite{Ricci2023} both show a shift in the continuum, but in the latter the broad lines disappear between epochs, a typical example of Changing Look quasars.
Finally, we made the conservative choice to exclude any quasar with a complex flux profile around the [OIII] line, e.g. coming from a strong Fe component.
We measured robust equivalent width ratios for 183 quasars and modeled them with a purely intrinsic and a combined intrinsic and lensing variability model.
Strong evidence in favour of the latter were found for 19 objects, which is sensible given that lensing is a rare phenomenon, but also reflects our conservative choices.

For these 19 quasars, the lensing variability depends mainly on the Einstein crossing time.
There is a degeneracy here between the velocity and mass of the lens: for the same crossing time we can have a more massive and faster lens or a less massive and slower one.
Despite our velocity model being a widely adopted one for microlensing~\cite{ChenIKai2023}, we plan to address this degeneracy in future work.
We have assumed a uniform population of lenses, but linking this to the galaxy distribution along the line of sight~\cite{WyitheTurner2002} can help us locate whether our detected lenses are within our Galaxy (self-lensing), other galaxy halos, or truly free-floating.
This can put tighter priors on the lens velocity and redshift, which would lead to better constraints on the mass.
Although the constraints on compact dark matter from~\cite{Zackrisson2007} remain largely unaffected by the choice of the spatial distribution of microlenses, it would still be valuable to take this into account while ranking our sample.

Our approach had a few conservative and simplistic aspects that we plan to improve on in the future.
We fitted the spectra only in a narrow range around [OIII] but we could extend our analysis to the entire spectrum, including multiple narrow lines and a differentially magnified continuum.
The latter would also lead to a measurement of the accretion disc temperature slope.
Any measurement of the Supermassive Black Hole mass of the quasars, which is currently a nuisance parameter that we marginalize over, would significantly improve our measurements.
We argue that this new approach could shed more light into dark matter as compact objects and it is worth pursuing, especially in connection to upcoming spectroscopic surveys like 4MOST.

\begin{acknowledgments}
The authors would like to thank Miguel Zumalac\'{a}rregui for very useful discussions of the context and importance of the results of our work and for commenting on a close-to-final version of the manuscript.
G.~V. has received funding from the European Union’s Horizon 2020 research and innovation programme under the Marie Sklodovska-Curie grant agreement No 897124.
F.~C. and J.~H.~H.~C. were supported by the Swiss National Science Foundation.
\end{acknowledgments}

\appendix

\section{Lens redshift prior}
\label{app:lens_redshift}
Because the lens is assumed to be a single compact object somewhere along the line of sight to a quasar, its redshift becomes an unknown, nuisance parameter that has to be marginalized over.
We can write:
\begin{eqnarray}
	p^{\rm LE} \left( \frac{\wo}{\wt}| \, \Delta t, z_{\rm s}, \mathbf{\zeta}, M \right) = \nonumber\\ 
	\int_0^{\zs} p^{\rm LE} \left( \frac{\wo}{\wt}| \, \Delta t, z_{\rm s}, \mathbf{\zeta}, M, \zl \right) p(\zl) \, \mathrm{d} \zl \; ,
    \label{eq:margin_zl}
\end{eqnarray}
\noindent where $p(\zl)$ is the probability distribution of the lens redshift.
In order to determine this distribution, we need to make assumptions on how the lenses are distributed in the Universe.

Our treatment assumes a lens population with a single mass that is non-evolving, i.e. has a constant comoving number density.
The number density of lenses as a function of redshift is given by:
\begin{equation}
	\label{eq:number_density}
	n(z) \equiv n(z|M,\Omega_{\rm l}) = (1+z)^3 n_{\rm 0} = \frac{3 \Omega_{\rm l} H_{\rm 0}^2 }{8 \pi G M} \, ,
\end{equation}
\noindent where $H_{\rm 0}$ is the Hubble constant, $G$ the gravitational constant, and $\Omega_{\rm l}$ the density parameter of the lenses.
The volume of a sphere whose edge is at $z$ is:
\begin{equation}
	\label{eq:volume}
	V(z) = 4 \pi D^2(z) \frac{c}{(1+z)H(z)} \, ,
\end{equation}
\noindent where $H(z)$ is the Hubble parameter and $c$ the speed of light.
Combining these two equations we get for the total number of lenses as a function of redshift:
\begin{equation}
	N(z) \equiv N(z|\Omega_{\rm l},M) \propto \frac{\Omega_{\rm l}}{M} \frac{(1+z)^2 D^2(z)}{H(z)} \, .
\end{equation}
\noindent From this last equation we can finally define $p(\zl)$ as:
\begin{equation}
	\label{eq:pzl}
	p(\zl|\zs) = \frac{N(\zl)}{\int_0^{\zs} N(\zl) \, \mathrm{d}\zl} \, ,
\end{equation}
\noindent where the dependence on $\Omega_{\rm l}$ and $M$ is cancelled out.

\section{Combined model}
\label{app:intrinsic_and_lensing}
For the variability model that combines lensing and intrinsic variability, the factor $\mu$ introduced in Eq.~(\ref{eq:ew_def_time}) is considered the product of the two independent random variables $\mu^{\mathrm{LE}}$ and $\mu^{\mathrm{IN}}$ whose distribution is given by:
\begin{equation}
	\label{eq:pmu_combined}
	p^{\rm CO}(\mu) = \int p^{\mathrm{LE}}\left(\mu^{\mathrm{LE}}\right) p^{\mathrm{IN}}\left(\frac{\mu}{\mu^{\mathrm{LE}}}\right) \frac{1}{\mu^{\mathrm{LE}}} \mathrm{d} \mu^{\mathrm{LE}},
\end{equation}
\noindent where $p^{\mathrm{LE}}$ is the probability due to lensing alone given in 
Eq.~(\ref{eq:prob_final_le}) and $p^{\mathrm{IN}}$ is the probability due to intrinsic variability alone, given in Eq.~(\ref{eq:prob_final_in}).
There is no difference whether we choose $\mu^{\mathrm{LE}}$ or $\mu^{\mathrm{IN}}$ as the independent variable in the integral.
An identical equation is valid for the initial factor $\mu_0$ at $t_0$.
Once we calculate the distributions $p^{\rm CO}_{\rm \mu}$ and $p^{\rm CO}_{\rm \mu_0}$ from Eq. (\ref{eq:pmu_combined}), we can replace them directly in Eq. (\ref{eq:ew_ratio_prob}) to obtain $p^{\rm CO} \left( \frac{\wo}{\wt} \big| \, \Delta t, M_{\rm i}, \zs, \mathbf{\zeta}, M \right)$.

\section{Bayesian factor}
\label{app:bayesian_factor}
We want to test which of the two variability models, Intrinsic or Combined, or hypotheses $H_{IN}$ and $H_{CO}$ respectively, is favoured by the data.
Before quantifying this, we explain the priors for the two models, which are based on the (Poisson) probability for no-lensing and lensing by a single object.

The average number of massive objects acting as lenses along the line of sight to some quasar is known as the lensing optical depth (see also eq. 106 in part 1 of~\cite{Schneider2006}):
\begin{eqnarray}
	\tau(\zs) \equiv \tau(\zs|M,\Omega_{\rm l}) = \nonumber \\
	\int_0^{\zs} \sigma(M,\zl,\zs) n(\zl|M,\Omega_{\rm l}) \frac{c}{(1+\zl) H(\zl)} \mathrm{d}\zl \, ,
    \label{eq:tau_theo}
\end{eqnarray}
\noindent where $n(\zl|M,\Omega_{\rm l})$ is the number density given in Eq. (\ref{eq:number_density}), and $\sigma(M,\zl,\zs)$ is the lensing cross-section.
The latter can be defined as the sky area around the lens where a source must be placed in order to produce multiple images.
The characteristic length for lensing is known as the Einstein radius, which for a point-mass lens of mass $M$ is given by:
\begin{equation}
	\label{eq:rein}
	\RE \equiv \RE(M,\zl,\zs) = \sqrt{ \frac{4GM}{c^2} \frac{D(\zl,\zs)D(\zs)}{D(\zl)} } \, ,
\end{equation}
\noindent where $D(\zl)$ and $D(\zs)$ are the angular diameter distances to the lens and source respectively, and $D(\zl,\zs)$ the angular diameter distance between lens and source.
Setting a threshold, $\mu_{\mathrm{thres}}$, for the lensing magnification, and requiring that $\mu \ge \mu_{\mathrm{thres}}$, translates into a maximum deflection angle, $\alpha_{\mathrm{thres}}$ and into a maximum source distance from the optical axis.
Here we use $\mu_{\mathrm{thres}} = 1.061$ that corresponds to $\alpha_{\mathrm{thres}} = 2 \RE$.
This results in a circular cross section equal to:
\begin{equation}
	\label{eq:cross_section}
	\sigma(M,\zl,\zs) = 4 \pi \RE^2 \, .
\end{equation}
\noindent Replacing Eqs. (\ref{eq:number_density}), (\ref{eq:rein}) and (\ref{eq:cross_section}) in (\ref{eq:tau_theo}) we get:
\begin{equation}
	\label{eq:tau_final}
	\tau(\zs|\Omega_{\rm l}) = \frac{6 \Omega_{\mathrm{l}}H_0^2}{c \, D(\zs)} \int_0^{\zs} \frac{(1+\zl)^2}{H(\zl)} D(\zl) D(\zl,\zs) \, \mathrm{d}\zl \, ,
\end{equation}
\noindent which does not depend on the individual lens mass, $M$, as it cancels out between the cross section and density terms, but only on the total mass in the form of compact objects through $\Omega_{\rm l}$.

\begin{figure}
	\centering
	\includegraphics[width=\columnwidth]{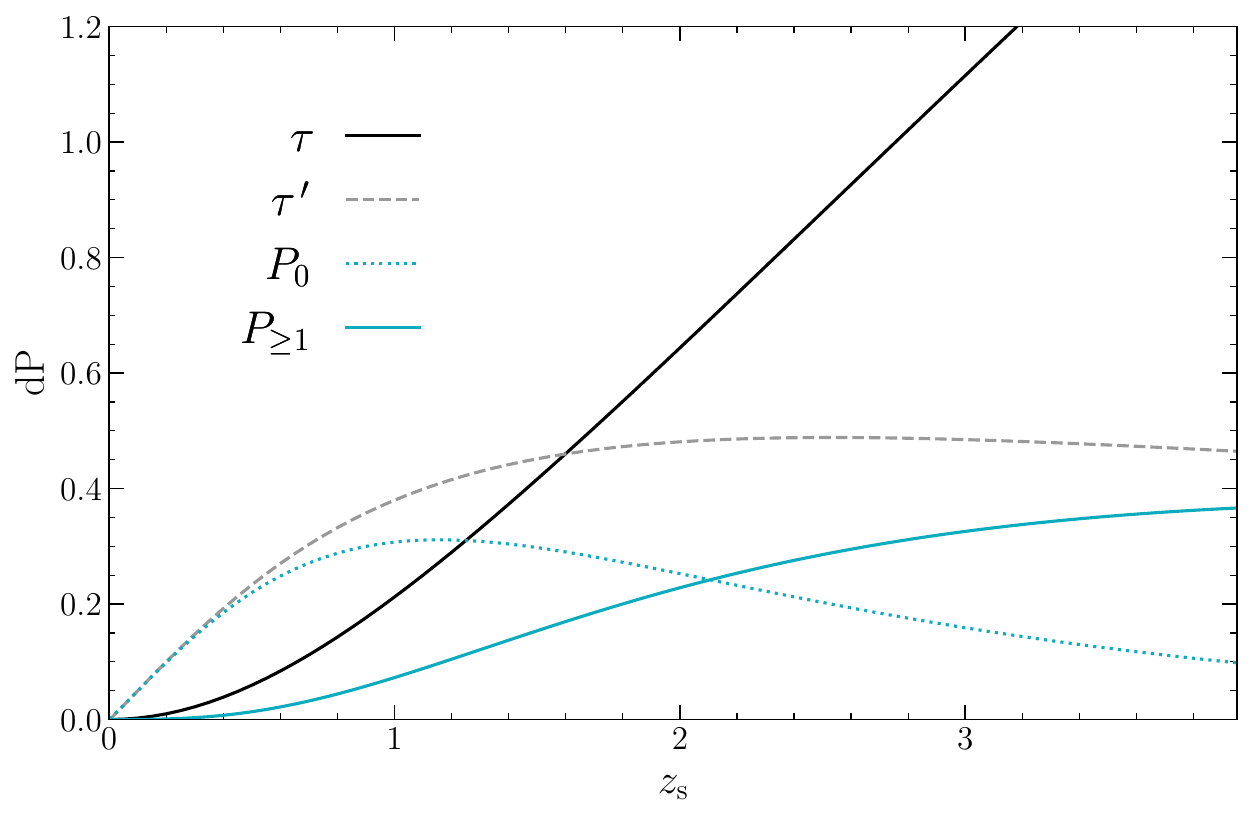}
	\caption{Probability density of lensing with a magnification threshold of 1.061. This is shown for none ($P_0$) or any number ($P_{\geq 1}$) of lenses as a function of redshift from Eq. (\ref{eq:poisson}). The optical depth from Eq.~(\ref{eq:tau_final}) and its derivative with respect to $\zs$ are shown in the same dimensionless scale. Here, we assumed all of dark matter to be in compact objects ($\Omega_{\mathrm{l}}=0.26$).}
	\label{fig:tau}
\end{figure}

We can now compute the Poisson probability of lensing taking place due to $k$ lenses, given $\Omega_{\mathrm{l}}$:
\begin{equation}
	\label{eq:poisson}
	P_{\rm k}(\zs|\Omega_{\mathrm{l}}) = \frac{\tau^{\rm k} e^{-\tau}}{\rm k!} \, \left| \frac{\mathrm{d}\tau}{\mathrm{d} \zs} \right| \quad \mathrm{with} \quad \tau \equiv \tau(\zs|\Omega_{\mathrm{l}}) \, ,
\end{equation}
We plot this probability as well as the optical depth in Fig.~\ref{fig:tau}.

To quantitatively test the Intrinsic and Combined model hypotheses, we use the Bayesian factor, $K$:
\begin{eqnarray}
	K\left( \frac{\wo}{\wt} \right) = \frac{P\left(H_{\rm CO}|\frac{\wo}{\wt}\right)}{P\left(H_{\rm IN}|\frac{\wo}{\wt}\right)} \nonumber \\
    = \frac{P\left(\frac{\wo}{\wt}|H_{\rm CO}\right)}{P\left(\frac{\wo}{\wt}|H_{\rm IN}\right)}\frac{P\left(H_{\rm CO}\right)}{P\left(H_{\rm IN}\right)} \nonumber\\    
    = \frac{ p^{\rm CO} \left( \frac{\wo}{\wt}| \, \Delta t, M_{\rm i}, z_{\rm s}, \mathbf{\zeta}, H_{\rm M} \right) P_{\rm 1}(z_{\rm s}|\Omega_{\rm l}) }{ p^{\rm IN} \left( \frac{\wo}{\wt}| \, \Delta t, M_{\rm i} \right)  P_{\rm 0}(z_{\rm s}|\Omega_{\rm l})} ,
    \label{eq:bayes_factor}
\end{eqnarray}
\noindent where the likelihoods $p^{\rm IN}$ and $p^{\rm CO}$ of the two models are given by Eqs. (\ref{eq:prob_final_in}) and (\ref{eq:pmu_combined}), and the priors for no-lensing and lensing by a single object result from setting ${\rm k}=0,1$ in Eq.~(\ref{eq:poisson}), which is only a function of the quasar redshift and the amount of matter in the form of compact objects, $\Omega_\mathrm{l}$.

\begin{figure}
	\centering
	\includegraphics[width=\columnwidth]{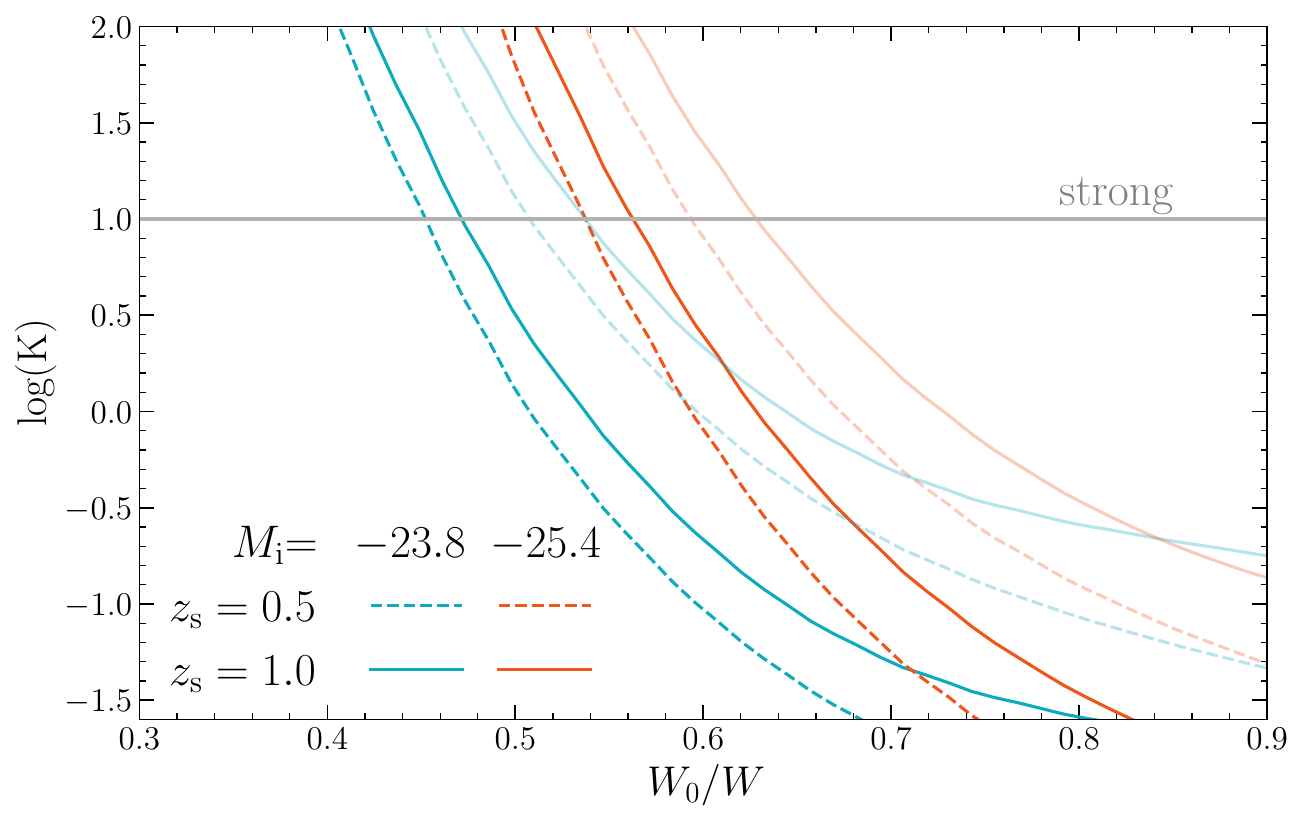}
	\caption{Bayesian factor calculated from Eq.~(\ref{eq:bayes_factor}) for some fiducial Combined and Intrinsic models. This is a function of the equivalent width ratio $\wo/\wt$ measured a decade after the reference time ($\Delta t=10$ years). The optical depth for lensing, and consequently $P_{\rm 0}$ and $P_{\rm 1}$, is calculated for $\Omega_{\rm l}/\Omega_{\rm DM} = 0.1 $ and $1.0$ for the heavy and faded lines respectively. The main effect of this is a shift by $\approx 0.1$ to lower $\wo/\wt$ of where $K$ intercepts the threshold for strong evidence in favour of the lensing hypothesis, $H_{\rm CO}$ ($\log K=1$ indicated by the horizontal line).}
	\label{fig:bayesian_factor}
\end{figure}

In Fig.~\ref{fig:bayesian_factor} we show the Bayesian factor $K$ for some fiducial Combined and Intrinsic models, with quasar redshifts and magnitudes in the $i$-band being representative of our sample, and a fixed lens mass of $0.3$ M$_{\odot}$.
It can be seen that primarily brighter quasars have a higher $K$ because the distribution of $\wo/\wt$ due to intrinsic variability becomes narrower in this case.
Higher redshift also leads to higher $K$ because the lensing optical depth increases, but this is a weaker effect.
For the example with $M_{\rm i}, \zs = (-25.4,1.0)$ (bright and distant quasar), and $\Omega_{\mathrm{l}}/\Omega_{\mathrm{DM}}=0.1 (1.0)$ we would have strong evidence ($\log K > 1$,~\cite{Jeffreys1998}) in favour of the lensing hypothesis if we observe $\wo/\wt \approx 0.6 (0.7)$ within a time interval of $\Delta t=10$ years.
The horizontal dashed lines in Fig. \ref{fig:bayesian_factor_sample} are purely due to the priors, which are the only terms that depend on $\Omega_{\mathrm{l}}$.
After manipulating the equations, these limits are directly proportional to $\log (\Omega_{\mathrm{l}}/\Omega_{\mathrm{DM}})$.

\section{Fitting the Combined model to the data}
\label{app:combined_fit}
If we denote the ratio of equivalent widths as $r=\wo/W$, then we can write its measured value for a single quasar as $\hat{r}_{\rm i} = r_{\rm i} + n_{\rm i}$, where $r_{\rm i}$ is its true value and $n_{\rm i}$ the measurement error.
We can now use Bayes theorem to write the probability of the lens mass given the measurement as:
\begin{eqnarray}
	p(M|\hat{r}_{\rm i}) \propto p(\hat{r}_{\rm i}|M) \, p(M) \nonumber \\
	= \int p(\hat{r}_{\rm i}|M,r_{\rm i}) \, p(r_{\rm i}) \, p(M) \, \mathrm{d} r_{\rm i} \nonumber \\ 
    = \int p(\hat{r}_{\rm i}|r_{\rm i}) \, p^{CO}(r_{\rm i}|M) \, p(M) \, \mathrm{d} r_{\rm i} \, .
	\label{eq:combined_prob}
\end{eqnarray}
\noindent The first term within the integral can be interpreted as the measurement error; assuming a gaussian distribution of the errors in the equivalent width measurements obtained from the spectra, it can be written as:
\begin{equation}
	\label{eq:data_error}
	p(\hat{r}_{\rm i}|r_{\rm i}) \propto \exp \left[ -\frac{1}{2} \left( \frac{\hat{r}_{\rm i}-r_{\rm i}}{\sigma_{\rm i}} \right)^2 \right] \, ,
\end{equation}
\noindent where $\sigma_{\rm i}$ is the error in the measured value of equivalent width ratio, $\hat{r}_{\rm i}$.
The second term is attributed to our model given by Eq. (\ref{eq:pmu_combined}), and requires the additional parameters $\zs, \mathbf{\zeta}, M_{\rm i}$ and $\Delta t$, which we omit here for clarity.
The third term can adopt any form to use as prior knowledge, here we chose the uninformative priors: uniform, logarithmic, and $M^{-1/2}(1-M)^{-1/2}$ (Jeffrey's prior) for the scaled mass between the adopted limits.
Finally, our measurements for a sample of $N$ quasars can be denoted by the vector $\mathbf{\hat{r}}$, and we can write:
\begin{equation}
	\label{eq:combined_prob_sample}
	p(M|\mathbf{\hat{r}}) \propto \prod_{\rm i}^{\rm N} p(M|\hat{r}_{\rm i}) \, .
\end{equation}
Assuming the same mass for all the lenses, the obtained mass limits at 99\% confidence for the three priors are:
\begin{itemize}
    \item Uniform: $0.014^{+0.005}_{-0.004}$ (shown in Fig. \ref{fig:combined_fit}),
    \item Logarithmic: $0.00032^{+0.00053}_{-0.00023}$,
    \item Jeffrey's: $0.0056^{+0.002}_{-0.0014}$.
\end{itemize}

\section{Supplementary material - final sample}
The properties of the final sample of 19 quasars are given in Table \ref{tab:sample}.
\begin{table}
	\caption{Sample of 19 quasars to which we fit the Combined model. The listed parameters are used to obtain the lens mass measurements shown in Fig.~\ref{fig:combined_fit}. The first five columns (coordinates, redshift, $i$-band magnitude, and $\Delta t$ in years between the two spectra) are obtained from SDSS, the ratio of equivalent widths $\wo/W$ is what we measure from the spectra, and the Bayesian factor $K$ is calculated for our two competing hypotheses: Intrinsic and Combined variability. The quasars are sorted by increasing $K$, which means a higher probability for lensing.}
	\centering
	\begin{tabular}{rrrrrrr} 
		RA & DEC & $\zs$ & $M_{\rm i}$ & $\Delta t$ & $\wo/W$ & log $K$ \\ 
		\hline
		228.469 & 49.5751 & 0.41 & -24.46 & 12.8 & 0.63$\pm$0.07 & 0.07  \\ 
		139.1059 & 47.2449 & 0.54 & -26.79 & 12.0 & 0.75$\pm$0.05 & 0.11 \\ 
		123.9919 & 50.209 & 0.6 & -26.43 & 14.0 & 0.74$\pm$0.04 & 0.19 \\ 
		40.668 & 0.9575 & 0.57 & -27.54 & 15.0 & 0.77$\pm$0.01 & 0.24 \\ 
		34.1981 & -1.2282 & 0.42 & -24.74 & 9.0 & 0.63$\pm$0.08 & 0.3 \\ 
		142.0393 & 38.5002 & 0.5 & -26.98 & 13.0 & 0.73$\pm$0.05 & 0.36 \\ 
		167.8405 & 48.3461 & 0.28 & -24.46 & 12.3 & 0.57$\pm$0.05 & 0.43 \\ 
		5.8848 & -1.2456 & 0.48 & -26.34 & 13.0 & 0.7$\pm$0.06 & 0.47 \\ 
		204.8016 & 53.9243 & 0.29 & -24.3 & 12.9 & 0.55$\pm$0.06 & 0.58 \\ 
		178.9767 & 44.9412 & 0.52 & -26.29 & 9.1 & 0.68$\pm$0.03 & 0.64 \\ 
		174.7028 & 57.2382 & 0.45 & -25.96 & 11.2 & 0.61$\pm$0.02 & 1.32 \\ 
		138.7143 & 42.3325 & 0.55 & -26.16 & 13.0 & 0.61$\pm$0.12 & 1.83 \\ 
		135.9208 & 39.2861 & 0.58 & -25.58 & 13.1 & 0.58$\pm$0.05 & 1.9 \\ 
		172.4855 & 36.822 & 0.4 & -23.88 & 10.9 & 0.45$\pm$0.1 & 2.18 \\ 
		123.6079 & 29.6877 & 0.37 & -24.43 & 8.0 & 0.47$\pm$0.12 & 2.49 \\ 
		141.7633 & 52.3879 & 0.6 & -27.08 & 13.2 & 0.61$\pm$0.05 & 2.78 \\ 
		215.825 & 48.5044 & 0.57 & -25.37 & 11.0 & 0.5$\pm$0.14 & 3.16 \\ 
		136.9387 & 53.406 & 0.71 & -26.73 & 13.8 & 0.55$\pm$0.21 & 3.97 \\ 
		120.9552 & 46.1695 & 0.51 & -25.6 & 13.2 & 0.42$\pm$0.09 & 5.9 \\ 
		\hline
	\end{tabular}
    \label{tab:sample}
\end{table}

\section{Supplementary material - equations}
We examine two basic and well-established variability models: 1) a purely intrinsic one (hereafter ``Intrinsic''), assumed to be a classic Damped Random Walk, and 2) intrinsic combined with lensing of a point-source by a point-mass lens along the line-of-sight  (hereafter ``Combined'').
Lensing by a single compact object is the most probable scenario for the redshift range of the quasars in our sample (see discussion on the optical depth in Appendix C) - a review of more complex magnification models for extragalactic microlensing can be found in~\cite{Bosca2022}.
The additional lensing variability results from the source having an effective velocity, i.e. the combination of the peculiar velocities of lens, source, and observer, and traversing the magnification field of the lens.
This leads to differential magnification of regions of the source of different sizes; here we assume that the continuum is in fact the point source that gets lensed while the line flux comes from a region too large to be affected by any compact object along the line-of-sight.

Each variability model provides an expression for $p_{\mu}(\mu|\Delta t,\mu_0)$ and $p_{\mathrm{\mu_0}}(\mu_0)$ that appear in Eq.~(\ref{eq:ew_ratio_prob}).
We emphasize here that each model has its own free parameters that can differ for any given quasar.
Hence, the left hand side of Eq. (\ref{eq:ew_ratio_prob}) (and accordingly $p_{\mathrm{\mu}}$ and $p_{\mathrm{\mu_0}}(\mu_0)$) explicitly becomes:
\begin{equation}
	\label{eq:ew_ratio_prob_in}
	p^{\rm IN} \left( \frac{\wo}{\wt} \big| \, \Delta t \right) \equiv p^{\rm IN} \left( \frac{\wo}{\wt}| \, \Delta t, M_{\rm i}, z_{\rm s} \right),
\end{equation}
\noindent for the Intrinsic model, where M$_{\rm i}$ is the absolute magnitude in the $i$-band and $z_{\rm s}$ is the quasar redshift, and for the Combined model:
\begin{equation}
	\label{eq:ew_ratio_prob_co}
	p^{\rm CO} \left( \frac{\wo}{\wt} \big| \, \Delta t \right) \equiv p^{\rm CO} \left( \frac{\wo}{\wt}| \, \Delta t, M_{\rm i}, z_{\rm s}, \mathbf{\zeta}, M \right),
\end{equation}
\noindent where M is the mass of the lens and $\mathbf{\zeta}$ the sky coordinates of the quasar.
For the lensing model, these parameters suffice to describe the effective velocity of the source across the magnification field of the lens that leads to lensing-induced variability.

We would like to draw an analogy of our probabilistic lensing model with standard lensing theory, in particular as presented in [see eq. 11.68 in~\cite{Schneider1992}].
Our approach can be interpreted as calculating the probability of a source at redshift $\zs$ to undergo a lensing event with a property $\wo/\wt$ at a given $\Delta t$, and the cross section can be interpreted as the area that the source can cover in $\Delta t$ under its effective velocity.

\subsection*{Variability due to lensing}
\label{app:lensing}
Lensing variability is due to one or more intervening massive and compact objects (lenses) at redshift $\zl$ passing along the line of sight to a quasar (source) at redshift $\zs$.
Different regions of the source with different physical sizes are then seen differentially magnified (see Fig. \ref{fig:schematic}).
This is captured by the factor $\mu(t)$ in Eq.~(\ref{eq:ew_def_time}), which is the ratio between the lensing magnifications for any given emission line and the underlying continuum flux.
The effect is time dependent due to the relative motions of the observer, the lens, and the source.
In this section we provide a quantitative description of this microlensing-induced variability.

\subsubsection*{Magnification}
We adopt a single and isolated point-mass lens.
Such microlenses do not produce any de-magnification as opposed to a collection of point-mass lenses, whose deflections are superimposed, i.e. the high optical depth microlensing regime~\cite{Paczynski1986,Kayser1986}, see also~\cite{VernardosISSI2024} for a recent review.
This means that any lensed region of the source can only be magnified, albeit by very small amounts for large angular separations from the lens (much larger than $\RE$, the Einstein radius of the system (see Eq.~\ref{eq:rein}).
Because the continuum region is more compact than the emission line regions, it is more prone to extreme magnifications.
Therefore, during a microlensing event we expect a stronger increase in the continuum flux compared to the lines.
This translates to a reduced observed line equivalent width in presence of microlensing than in its absence.
In other words, we always have $\mu(\Delta t,\mu_0) > 1$.

In the current quasar structure paradigm, the most compact region is the continuum accretion disk, followed by the Broad Line Region (BLR) and then the Narrow Line Region (NLR), e.g.~\cite{Antonucci1993}, see also fig. 3 of~\cite{VernardosISSI2024}.
While it is known that microlensing can affect the BLR~\cite{Sluse2007, Sluse2012}, it leaves the much larger NLR unchanged~\cite{Moustakas2003, Nierenberg2017}.
We therefore focus here on the relative effect of microlensing on the continuum and the NLR.
This simplifies the expression of the $\mu(\Delta t,\mu_0)$ factor, as the ratio of the line and continuum magnifications can then be replaced by the magnification of the continuum alone, assumed to be a point-like source.
We note that any finite-size source effect, i.e. the size of the continuum being a considerable fraction of the Einstein radius of the lens, would mostly lead to less extreme magnification values.
As will be shown below, our analysis is not affected by extremely high magnification values, which means that our point-source assumption is justified.

For a point-mass lens in the absence of shear\footnote{Shear is an effect of the environment around the lens, e.g. induced by the tidal shearing effect of galaxies lying along the line of sight. Our assumption for no shear is in line with assuming a uniform spatial distribution of lenses, for which tidal effects are expected to be small or cancel out (there is no dominant mass distribution along the line of sight). For a lens number density that is associated with the density of galaxies along the line of sight see~\cite{Zackrisson2007}.} only two lensed images of the source are formed.
The total magnification for both these images is:
\begin{equation}
	\label{eq:mag}
	\mu \equiv \mu(\rs) = \frac{\rs^2+2}{\rs \sqrt{\rs^2 + 4}} \, ,
\end{equation}
\noindent where $\rs$ is the angular separation between the optical axis, centered on the lens, and the source, seen in projection on the plane of the sky (see Fig.~\ref{fig:schematic}).
Knowing the probability density of a random variable, $X$, we can derive the one for $Y=g(X)$ by performing a change of variables and taking into account the fact that the probability contained in an infinitesimal area must be conserved:
\begin{equation}
	\label{eq:change_of_variables}
	p_{\mathrm{Y}}(y) = p_{\mathrm{X}}\left( \, g^{-1}(y) \, \right) \left| \frac{\mathrm{d} \, g^{-1}(y)}{\mathrm{d} y} \right|,
\end{equation}
\noindent where $g^{-1}$ is the inverse of $g$.
In our case, assuming $\rs$ to be a random variable we can get the probability density of $\mu$ through inverting Eq.~(\ref{eq:mag}) to get $\rs \equiv \rs(\mu)$:
\begin{equation}
	\label{eq:pmu_lensing}
	p_{\mathrm{\mu}}(\mu) \equiv p_{\mathrm{\mu}}(\mu | \Delta t, \mu_{\mathrm{0}}) = p_{\mathrm{\rs}} \left( \rs(\mu) \right) \frac{1}{\rs(\mu)\left( \mu^2 -1\right)^{3/2}},
\end{equation}
\noindent where $\prs$ is the probability for a displacement $\rs$ on the source plane (derived in detail below).
We note that this probability is conditioned on $\Delta t$ and $\mu_{\mathrm{0}}$.
Equation~(\ref{eq:mag}) and consequently (\ref{eq:pmu_lensing}) are derived analytically, but any extensions to our lensing model, e.g. finite-sized continuum region and lensed emission line flux, would require a numerical equivalent.

The probability distribution of $\mu_0$ is obtained by assuming $\rs$ to be a uniform random variable, i.e. unconditioned on $\Delta t$ and $\mu_0$, between 0 and an upper limit corresponding to a lower magnification threshold, $\mu_{\rm thres}$:
\begin{equation}
	\label{eq:pmu0_lensing}
	p_{\mathrm{\mu_0}}(\mu_0) = \frac{\sqrt{\mu_{\rm thres}^2 - 1}}{\mu_{\rm thres}-\sqrt{\mu_{\rm thres}^2 - 1}} \times \frac{1}{(\mu_0^2-1)^{3/2}}.
\end{equation}
\noindent The requirement for such a threshold will be evident in the calculation of the optical depth later on.

\subsubsection*{Displacement in the source plane}
\label{sec:displacement}
The magnification given in Eq.~(\ref{eq:mag}) is a function of the displacement, $\rs$, in the source plane, which can be written as the sum of two vectors (as indicated in Fig.~\ref{fig:schematic}):
\begin{equation}
	\label{eq:rs_sum}
	\rs \equiv |\mathbf{\rs}| = |\mathbf{r}_0 + \mathbf{r}_\mathrm{\upsilon}| = \left( r_0^2 + \rv^2 - 2 \, r_0 \, \rv \, \cos \phi_{\mathrm{\upsilon}} \right)^{1/2}.
\end{equation}
\noindent The first one, $r_0$, is a constant vector indicating the location of the source at the time of the first observation, $t_0$, corresponding to a magnification $\mu_0$.
The second, $\rv$, is the displacement at time $t$ due to the relative motions between observer, lens, and source.
The angle $\phi_{\mathrm{\upsilon}}$ above is between vectors $r_0$ and $\rv$.
The magnitude of vector $\mathbf{\rv}$ is a random variable whose time-dependent probability distribution, $p(\rv)$, will be derived below.
Thus, $\rs$ is distributed according to:
\begin{equation}
	\label{eq:prs_start}
	p_{\mathrm{\rs}}(\rs) = \int_{-\infty}^{+\infty} p_{\mathrm{\phi_{\upsilon},\rv}}\left(\phi_{\mathrm{\upsilon}} , \rv \right) \left| \frac{\partial \phi_{\mathrm{\upsilon}}}{\partial \rs} \right| \mathrm{d} \rv \, ,
\end{equation}
\noindent where:
\begin{equation}
	\label{eq:phi_function}
	\phi_{\mathrm{\upsilon}} \equiv \phi_{\mathrm{\upsilon}}(\rs,\rv) = \cos^{-1} \left( \frac{r_0^2 + \rv^2 - \rs^2}{2\rv r_0} \right) \, ,
\end{equation}
\noindent and hence,
\begin{equation}
	\label{eq:phi_derivative}
	\frac{\partial \phi_{\mathrm{\upsilon}}}{\partial \rs} = 2\rs \Big\{ \big[\rv^2-(r_0 - \rs)^2\big]\big[(r_0+\rs)^2 - \rv^2\big]\Big\}^{-1/2}.
\end{equation}
\noindent However, $\rv$ and $\phi_{\mathrm{\upsilon}}$ are independent variables, which leads to:
\begin{equation}
	\label{eq:split_joint}
	p_{\mathrm{\phi_{\upsilon},\rv}}\left(\phi_{\mathrm{\upsilon}} , \rv \right) = p_{\mathrm{\phi_{\upsilon}}}(\phi_{\mathrm{\upsilon}}) p_{\mathrm{\rv}}(\rv) = \frac{1}{\pi}p_{\mathrm{\rv}}(\rv) \, ,
\end{equation}
\noindent where we used the fact that $\phi_{\mathrm{\upsilon}}$ has a uniform distribution between 0 and $\pi$\footnote{This is not entirely true because, as we will see in the next section, the relative velocity has a fixed component with respect to the Cosmic Microwave Background, which introduces a preferred direction. However, it turns out that this velocity component has a very small contribution to $\rv$ (see Fig. \ref{fig:prob_rv}) and can thus be safely ignored.}.
Finally, combining Eqs. (\ref{eq:prs_start}), (\ref{eq:phi_derivative}), and (\ref{eq:split_joint}), we obtain:
\begin{eqnarray}
	p_{\mathrm{\rs}}(\rs) = \nonumber \\ 
	\frac{2\rs}{\pi} \int_{r_0-\rs}^{r_0+\rs} \frac{p_{\mathrm{\rv}}(\rv) \, \mathrm{d} \rv}{\Big\{ \big[\rv^2-(r_0 - \rs)^2\big]\big[(r_0+\rs)^2 - \rv^2\big]\Big\}^{1/2}}  \, , \nonumber \\
    \label{eq:prs_final}
\end{eqnarray}
\noindent where the integration limits are defined by the requirement that the quantity within the square root be larger than zero.

\subsubsection*{Probability distribution for $r_{\mathrm{\upsilon}}$}
\label{sec:prob_rv}
The net effect of the relative motions between observer, lens, and source, can be described by an effective velocity on the source plane~\cite{Kayser1986}.
The different velocity components at work are the velocity of the observer and the peculiar velocities of the lens and the source. 
These are the same velocity components used in quasar microlensing studies~\cite{Neira2020}.

The velocity of the observer on the plane of the sky is defined as the transverse velocity component with respect to the three-dimensional velocity due to the Cosmic Microwave Background (CMB) dipole~\cite{Kogut1993}:
\begin{equation}
	\label{eq:vcmb_theo}
	\mathbf{\upsilon}_{\mathrm{trans}} = \mathbf{\upsilon}_{\mathrm{dip}} - (\mathbf{\upsilon}_{\mathrm{dip}} \cdot \hat{z}) \hat{z} \, ,
\end{equation}
\noindent where $\hat{z}$ is the unit vector along the line of sight, and $\mathbf{\upsilon}_{\mathrm{dip}}$, our velocity with respect to the CMB, is a known function of sky coordinates [see fig.~1 of~\cite{Neira2020}].
We define the magnitude of this vector, projected onto the source plane as:
\begin{equation}
	\label{eq:vcmb}
	\upsilon_{\mathrm{CMB}} \equiv \upsilon_{\mathrm{CMB}}(\mathbf{\zeta},\zl,\zs) = \frac{|\mathbf{\upsilon}_{\mathrm{trans}}|}{1+\zl} \frac{D(\zl,\zs)}{D(\zl)} \, ,
\end{equation}
\noindent where $\zl$ and $\zs$ are the redshifts of the lens and source, $D(\zl)$ the angular diameter distance to the lens, and $D(\zl,\zs)$ the angular diameter distance between lens and source.

The peculiar velocities of the lens and the source are random vectors, whose magnitude and direction have a normal and uniform distribution respectively - here we assume that the compact lenses, regardless of their true nature, have the same peculiar velocity distribution as galaxies, whether they happen to lie or not within galaxies. The one-dimensional normal distribution of these peculiar velocities has a zero mean and a redshift dependent standard deviation given by:
\begin{equation}
	\label{eq:sigma_pec}
	\sigma_{\mathrm{pec}}(z) = \frac{\sigma_{\mathrm{pec}}(0)}{\sqrt{1+z}}\frac{f(z)}{f(0)} \, ,
\end{equation}
\noindent where for the growth factor, $f(z)$, we use the approximation by~\cite{Lahav1991}, i.e. $f(z) \approx \Omega^{0.6}(z)$, and we assume $\sigma_{\mathrm{pec}}(0) = 235$ km/s~\cite{Kochanek2004,Blackburne2009}.
Because the peculiar velocities of the lenses and sources are independent random normal variables, their sum, after projecting onto the source plane, will also be normally distributed with its standard deviation being the quadratic sum of the individual distributions:
\begin{equation}
	\label{eq:combined_sigma_pec}
	\sigma_{\mathrm{\upsilon}}^2 = \left[ \frac{\sigma_{\mathrm{pec}}(\zl)}{1+\zl} \frac{D(\zs)}{D(\zl)} \right]^2 + \left[ \frac{\sigma_{\mathrm{pec}}(\zs)}{1+\zs} \right]^2 \, .
\end{equation}

Because we are using Eq. (\ref{eq:mag}) to find the magnification, we need the displacement on the source plane in units of the Einstein radius projected on the source plane, $\RE$, which for a point-mass lens is given by:
\begin{equation}
	\label{eq:rein}
	\RE \equiv \RE(M,\zl,\zs) = \sqrt{ \frac{4GM}{c^2} \frac{D(\zl,\zs)D(\zs)}{D(\zl)} } \, ,
\end{equation}
\noindent where $G$ is the gravitational constant.
Therefore, for a given time interval, $\Delta t$, we define:
\begin{equation}
	\label{eq:v_to_r_1}
	r_{\mathrm{c}} \equiv r_{\mathrm{c}}(\mathbf{\zeta},\Delta t,\zl,\zs,M) = \frac{\upsilon_{\mathrm{CMB}} \, \Delta t}{\RE}
\end{equation}
\noindent and
\begin{equation}
	\label{eq:v_to_r_2}
	\sigma \equiv \sigma(\Delta t,\zl,\zs,M) = \frac{\sigma_{\mathrm{\upsilon}} \, \Delta t}{\RE} \, ,
\end{equation}
\noindent where we used the fact that the product of a normal variable (the combined projected peculiar velocities) with a constant (the factor $\Delta t/\RE$) is also normally distributed.
The displacement vector during a given time interval, $\mathbf{r}_{\mathrm{\upsilon}}$, is, therefore, the sum of the constant vector $r_{\mathrm{c}}$ due to the projected transverse velocity with respect to the CMB dipole and a random, normally-distributed one due to the combination of the projected peculiar velocities.
The magnitude of such a vector, $\rv$, follows the Rice distribution:
\begin{eqnarray}
	p(\rv| r_{\mathrm{c}},\sigma) \equiv p(\rv| \mathbf{\zeta},\Delta t,\zl,\zs,M) = \nonumber \\
	\frac{\rv}{\sigma^2} I_0 \left( \frac{\rv r_{\mathrm{c}}}{\sigma^2} \right) \mathrm{exp}\left( - \frac{\rv^2 + r_{\mathrm{c}}^2}{2\sigma^2} \right) \, ,
    \label{eq:rice}
\end{eqnarray}
\noindent where $I_0$ is the zeroth order modified Bessel function of the first kind.
This distribution has an implicit dependence on time through Eqs. (\ref{eq:v_to_r_1}) and (\ref{eq:v_to_r_2}), on the lens redshift and mass (the latter through $\RE$), and on the source redshift and sky coordinates, $\mathbf{\zeta}$ (the latter through $\upsilon_{\mathrm{CMB}}$).

\begin{figure}
	\centering
	\includegraphics[width=\columnwidth]{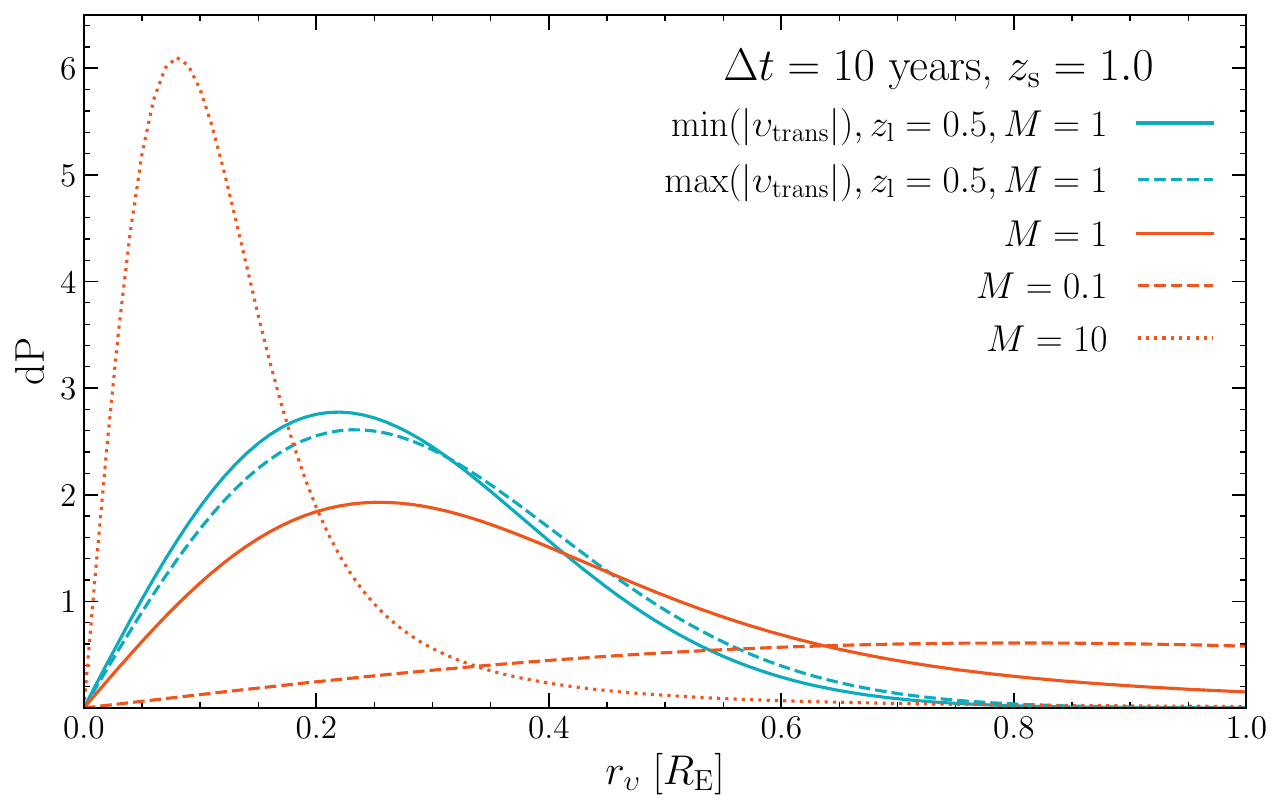}
	\caption{Effect of different parameters on the probability density of the displacement, $\rv$. This is in units of the Einstein radius, $\RE$, of a hypothetical source at $\zs = 1$ over a 10 year period. For a fixed lens mass and redshift we show two realizations (blue lines), one with the minimum (solid) and one with the maximum (dashed) transverse velocity allowed by the CMB, $\upsilon_{\mathrm{trans}}$ in Eq.~(\ref{eq:vcmb_theo}). These are described as Rice distributions from Eq.~(\ref{eq:rice}). Marginalizing over $\zl$, we show the effect of the lens mass (red lines), assumed to have a fixed value (in units of M$_{\odot}$).}
	\label{fig:prob_rv}
\end{figure}

In Fig.~\ref{fig:prob_rv}, we show a few examples of distributions of $\rv$ corresponding to different parameters.
Notably, the velocity due to the CMB barely affects the shape of the distribution given by Eq. (\ref{eq:rice}). This happens because $r_{\mathrm{c}}$ is always much smaller than $\sigma$ (Eq. \ref{eq:v_to_r_2}), which also explains why the shape of the curves in Fig.~\ref{fig:prob_rv} looks more like a Rayleigh distribution rather than a Gaussian - the asymptotic forms of the Rice distribution for very small and very large values of $r_{\mathrm{c}}$ compared to $\sigma$ respectively.
However, the mass of the lens plays a significant role in determining the resulting values of $\rv$: for small masses, $M<1$ M$_\odot$, the distribution gets narrow and peaks below 20 per cent of $\RE$, while for larger ones it extends to a few $\RE$.
This is because $\RE$, which is proportional to the square root of the mass, is in the denominator of Eqs.~(\ref{eq:v_to_r_1}) and ~(\ref{eq:v_to_r_2}).

\subsection*{Intrinsic variability}
\label{app:intrinsic}
The accretion mechanism that produces the continuum flux in quasars is known to be intrinsically variable.
The subsequent variability propagates outwards to the broad line region and then to the narrow line region.
Because here we examine the equivalent widths of narrow lines, which originate from a region orders of magnitude larger than the continuum, we can safely assume that the factor $\mu(t)$ in Eq.~(\ref{eq:ew_def_time}) is due to the variability of the continuum only.
This means that we can write the continuum flux as $F_C(t) = \mu(t) \langle F_C \rangle$, resulting in:
\begin{equation}
	\label{eq:mu_intrinsic}
	\mu(\Delta t,\mu_0) = 10^{0.4*[\langle M \rangle - M(\Delta t,\mu_0)]} \, ,
\end{equation}
\noindent where $M$ is the continuum flux in magnitudes (not to be confused with mass).
If we also define:
\begin{equation}
	\label{eq:X}
	X=0.4[\langle M \rangle - M(\Delta t,\mu_0)],
\end{equation}
\noindent then:
\begin{equation}
	\label{eq:pmu_intrinsic}
	p_{\mathrm{\mu}}(\mu|\Delta t,\mu_0) = p_{\mathrm{X}}(\mathrm{log} \mu|\Delta t,\mu_0) \frac{1}{\mathrm{ln}10} \frac{1}{\mu}.
\end{equation}
\noindent The resulting distribution of the factor $\mu$ depends on the description of $M$ as a function of time.

A very successful model for continuum variability in quasars has been the stochastic Damped Random Walk model (DRW).
In this work, we adopt this model, following~\cite{MacLeod2010}, who fit the DRW description to the SDSS stripe 82 quasars.
The DRW model states that $M(\Delta t)$ at some time interval $\Delta t$ after an initial observation $M_0$ is drawn from a normal distribution whose mean and variance are functions of two characteristic time and amplitude scale parameters, $\tau$ and $SF$, and of time.
The dependence of $X$ on $M$ is linear, which means that it also follows a normal distribution with mean and variance given from a slightly modified version of Eq.~(5) in~\cite{MacLeod2010}:
\begin{eqnarray}
	\label{eq:X_mean_var}
	E_{\mathrm{X}} &=&  e^{-\Delta t/\tau}\times \mathrm{log}\mu_0 \\
	V_{\mathrm{X}} &=&  \frac{1}{2} (0.4 \times SF)^2 (1-e^{-2\Delta t/\tau}).
\end{eqnarray}
\noindent These authors also provide a fit for the characteristic time scale, $\tau$, and scale, $SF$,  as a function of wavelength, for the $K$-corrected absolute magnitude in the $i$-band, $M_{\mathrm{i}}$, and for the black hole mass, $M_{\mathrm{BH}}$ [eq. 7 in~\cite{MacLeod2010}].
In addition, the black hole mass has a distribution parametrised by $M_{\mathrm{i}}$, so that:
\begin{eqnarray}
	p_{\mathrm{X}}(x|\Delta t,\mu_0,M_{\rm i}) = \nonumber \\
	\int p_{\mathrm{X}}(x|\Delta t,\mu_0,\mathrm{log}M_{\mathrm{BH}}) p_{\mathrm{M_{BH}}}(\mathrm{log}M_{\mathrm{BH}}|M_{\rm i}) \, d \mathrm{log}M_{\mathrm{BH}}, \nonumber \\
    \label{eq:mbh_marginalization}
\end{eqnarray}
\noindent where $p_{\mathrm{M_{BH}}}$ is a normal distribution whose mean and standard deviation are equal to $2.0-0.27M_{\mathrm{i}}$ and $0.58+0.011M_{\mathrm{i}}$ [eq. 11 in~\cite{MacLeod2010}].
This equation completely defines the right hand side of Eq.~(\ref{eq:pmu_intrinsic}), where apart from the values of $\Delta t$ and $\mu_0$, the only other required parameters are the wavelength of observation and the $K$-corrected absolute magnitude in the $i$-band.
Although in principle the black hole masses can be estimated from the H$\beta$ line width from the actual quasar spectra, we treat them as distributions in the context of this work so that our calculation remains general enough to be applied on populations of quasars rather than specific objects. 

Finally, the distribution of $\mu_0$ at the onset of the observations does not depend on time and can be derived from Eq.~(\ref{eq:mbh_marginalization}), in the limit of infinitely long time scales, $\Delta t$, in $p_{\mathrm{X}}$.
This also makes it independent of any initial value $\mu_0$ and leads to a normal distribution with zero mean and variance equal to $0.08 \times SF^2$.


\end{document}